**Title**: A *Causal Roadmap* for Hybrid Randomized and Real-World Data Designs: Case Study of Semaglutide and Cardiovascular Outcomes

**Authors and Affiliations:**

Dang LE[1], Fong E[2], Tarp JM[2], Clemmensen KKB[2], Ravn H[2], Kvist K[2], Buse JB[3], van der Laan M*[1], Petersen M*[1]

*Co-Senior Authors

[1]Department of Biostatistics, University of California, Berkeley, CA, USA

[2]Novo Nordisk, Søborg, Denmark

[3]Division of Endocrinology, Department of Medicine, University of North Carolina, Chapel Hill, NC, USA

**Corresponding Author:**

Lauren Eyler Dang

Division of Biostatistics

University of California, Berkeley

2121 Berkeley Way, Room 5302

Berkeley, CA 94720

(510) 642-3241

Lauren.eyler@berkeley.edu

**Conflict of Interest Statement: LED** reports tuition and stipend support from a philanthropic gift from the Novo Nordisk corporation to the University of California, Berkeley to support the Joint Initiative for Causal Inference. **EF, JMT, KKBC, HR, and KK** are full-time employees of Novo Nordisk A/S and own stocks in Novo Nordisk A/S. **KKBC** is also affiliated with Steno Diabetes Center Copenhagen, Denmark. **JBB** reports contracted fees and travel support for contracted activities for consulting work paid to the University of North Carolina by Novo Nordisk; grant support by Dexcom, NovaTarg, Novo Nordisk, Sanofi, Tolerion and vTv Therapeutics; personal compensation for consultation from Alkahest, Altimmune, Anji, AstraZeneca, Bayer, Biomea Fusion Inc, Boehringer-Ingelheim, CeQur, Cirius Therapeutics Inc, Corcept Therapeutics, Eli Lilly, Fortress Biotech, GentiBio, Glycadia, Glyscend, Janssen, MannKind, Mellitus Health, Moderna, Pendulum Therapeutics, Praetego, Sanofi, Stability Health, Terns Inc, Valo and Zealand Pharma; stock/options in Glyscend, Mellitus Health, Pendulum Therapeutics, PhaseBio, Praetego, and Stability Health; and board membership of the Association of Clinical and Translational Science. **MvdL** reports that he is a co-founder of the statistical software start-up company TLrevolution, Inc. **MvdL and MP** report personal compensation for consultation from Novo Nordisk.

**Abstract**

**Introduction:** Increasing interest in real-world evidence has fueled the development of study designs incorporating real-world data (RWD). Using the *Causal Roadmap*, we specify three designs to evaluate the difference in risk of major adverse cardiovascular events (MACE) with oral semaglutide versus standard-of-care: **1)** the actual sequence of non-inferiority and superiority randomized controlled trials (RCTs), **2)** a single RCT, and **3)** a hybrid randomized-external data study.

**Methods:** The hybrid design considers integration of the PIONEER 6 RCT with RWD controls using the experiment-selector cross-validated targeted maximum likelihood estimator. We evaluate 95% confidence interval coverage, power, and average patient-time during which participants would be precluded from receiving a glucagon-like peptide-1 receptor agonist (GLP1-RA) for each design using simulations. Finally, we estimate the effect of oral semaglutide on MACE for the hybrid PIONEER 6-RWD analysis.

**Results:** In simulations, **Designs 1 and 2** performed similarly. The tradeoff between decreased coverage and patient-time without the possibility of a GLP1-RA for **Designs 1 and 3** depended on the simulated bias. In real data analysis using **Design 3**, external controls were integrated in 84% of cross-validation folds, resulting in an estimated risk difference of -1.53%-points (95% CI -2.75%-points to -0.30%-points).

**Conclusions:** The *Causal Roadmap* helps investigators to minimize potential bias in studies using RWD and to quantify tradeoffs between study designs. The simulation results help to interpret the level of evidence provided by the real data analysis in support of the superiority of oral semaglutide versus standard-of-care for cardiovascular risk reduction.

**Introduction**

Semaglutide, a glucagon-like peptide-1 receptor agonist (GLP-1RA) developed as an antihyperglycemic agent, has been shown to improve multiple health outcomes for patients with type 2 diabetes mellitus (T2DM). In the SUSTAIN series of randomized controlled trials (RCTs), injectable semaglutide decreased glycated hemoglobin (HbA1c), body weight, and systolic blood pressure compared to placebo[1], sitagliptin[2], exenatide ER[3], and insulin glargine[4]. In the SUSTAIN 6 trial, injectable semaglutide decreased rates of major adverse cardiovascular events (MACE: defined as death from cardiovascular causes or nonfatal stroke or myocardial infarction (MI)) compared to placebo in patients with high cardiovascular (CV) risk, with an estimated hazard ratio (HR) of 0.74 (95% confidence interval (CI), 0.58 to 0.95)[5]. As a result, the United States Food and Drug Administration (FDA) approved injectable semaglutide for adults with T2DM to improve glycemic control and reduce cardiovascular risk in patients with cardiovascular disease.

Oral semaglutide was subsequently developed and shown, in the PIONEER series of RCTs, to decrease HbA1c compared to placebo[6], empagliflozin[7], and sitagliptin[8], and body weight compared to placebo[6], sitagliptin[8], and liraglutide[9]. To satisfy a pre-approval regulatory requirement for demonstrating cardiovascular safety, the PIONEER 6 RCT was designed to evaluate non-inferiority of oral semaglutide versus placebo[10]. For the primary outcome of MACE, the estimated HR was 0.79 (95% CI, 0.57 - 1.11), a result that was statistically significant for non-inferiority[10]. The evidence obtained through the PIONEER and SUSTAIN trials was not deemed sufficient for FDA approval of oral semaglutide for the secondary indication of cardiovascular risk reduction, prompting initiation of the ongoing SOUL RCT that has enrolled over 9,500 participants[11].

A superiority RCT is a standard choice for evaluating the effect of interest, yet RCTs may also have downsides. For example, clinicians treating placebo arm patients are directed not to prescribe medications of the same class as the active treatment under investigation[11–13]. Yet in 2019, a joint statement by the American Diabetes Association and the European Association for the Study of Diabetes emphasized that "for patients with type 2 diabetes and established atherosclerotic CV disease … where MACE is the gravest threat, the level of evidence for MACE benefit is greatest for GLP-1 receptor agonists"[14]. Although none of the trial participants were taking a GLP-1RA at baseline[11], would it not be better for the participants in the placebo arm of SOUL if they were allowed to start a GLP-1RA? This question led us to ask whether alternate trial designs incorporating real-world data (RWD) could decrease the amount of participant-time during which commencement of a GLP1-RA is precluded.

One such design, the hybrid RCT-external data study, has been increasingly utilized for estimating effects of medications for rare diseases[15] and/or pediatric[16] drug approvals when running an adequately powered RCT may not be feasible. Including external data may improve power but may also increase bias. An important subset of hybrid designs, including that considered here[17], estimate the bias that would be introduced by including non-randomized data in the analysis in order to decide whether to estimate the effect of interest based solely on the RCT, based on the pooled RCT and RWD, or based on some weighted combination of the two[17–22].

In this case study, we evaluate the utility of integrating an external control arm for a common disease, T2DM, with RCT data to estimate benefits for a secondary indication. In this context, an RCT is possible but has potential disadvantages for patient care, while a hybrid RCT-RWD analysis could decrease the amount of time patients are precluded from receiving a GLP1-RA yet raises questions about whether the resulting effect estimate is causal or merely associative. To evaluate these potential tradeoffs, we demonstrate use of the *Causal Roadmap*[23–25] – a structured process that helps investigators to pre-specify analytic study designs incorporating RWD – to compare different designs that could be used to estimate the effect of oral semaglutide on cardiovascular outcomes. An overview of the *Causal Roadmap* may be found in the companion paper[26]. Specifically, we compare a traditional program of RCTs to a hybrid randomized-RWD study integrating data from the PIONEER 6 non-inferiority trial with external controls from Optum's de-identified Clinformatics® Data Mart Database (CDM) (2007-2022) through simulations that closely mimic these true experiments. We then present results of the real hybrid analysis of PIONEER 6 and CDM for the estimated difference in the risk of a combined outcome of first MI, stroke, or all-cause death with oral semaglutide versus standard-of-care (without a GLP1-RA).

**Materials and Methods**

Table 1 describes design and analysis plans for three potential study designs for evaluating this question, using the list of *Causal Roadmap* steps found in the companion article[26]; additional detail is provided in the text and appendices. As visually depicted in Figure 1, **Design 1** is based on what truly occurred – a non-inferiority trial was run to demonstrate cardiovascular safety of oral semaglutide (PIONEER 6), after which, due to results that were promising but non-significant for superiority, a superiority trial was initiated. **Design 2** considers the hypothetical scenario in which only the superiority RCT is run, as might have occurred if superiority had been expected from the start. **Design 3** is a hybrid RCT-RWD study in which first a non-inferiority trial potentially augmented with extra RWD controls is run, and the follow-up superiority RCT is only initiated if the hybrid design does not reject the null hypothesis.

**Table 1: *Causal Roadmap* Steps for Specification of Study Designs 1-3**

| Roadmap Step | Designs 1-2 (RCT Only) | Design 3 (RCT + RWD) |
|---|---|---|
| 1a. Causal Question/ Causal Estimand | What would the difference in risk of MACE$^§$ (defined as death from any cause, nonfatal MI, or nonfatal stroke) within one year be if all patients in a population consistent with the PIONEER 6 inclusion/exclusion criteria and timeframe[10], and with similar healthcare engagement, were prescribed oral semaglutide plus standard-of-care compared to if all patients were prescribed standard-of-care alone, and if censoring had been prevented for all patients? See Appendix 1 for the mathematical representation of the causal estimand for each study design. | |
| 1b. Causal Model | Knowledge about potential shared causes of treatment, censoring, MACE, and participation in the RCT vs RWD, as well as possible causal relations between these variables depicted in Figure 2. | |
| 2. Observed Data | Potential data sources: Pioneer 6 RCT, SOUL RCT | Potential data sources: Pioneer 6 RCT, Optum CDM control arm, SOUL RCT |
| 3. Assess Identification and 4. Specify Statistical Estimand | Identification highly likely (non-administrative censoring in PIONEER 6 only 0.3%)<br><br>Statistical Estimand: Risk difference between treatment and control arms of the trial. | Plausible, though uncertain, that causal gap[§§] would be small (see Step 6).<br><br>Statistical Estimand: Adjusted risk difference between treatment and control arms, standardized to the covariate distribution in the target population. |
| 5. Statistical Model and Estimator | Statistical Model: Semi-parametric statistical model (incorporating knowledge that treatment was randomized).<br><br>Estimator: Unadjusted difference in risk between arms. | Statistical Model: Semi-parametric statistical model (incorporating knowledge that treatment in the RCT was randomized).<br><br>Estimator: Experiment-Selector CV-TMLE |
| 6. Sensitivity Analysis | None given that causal identification assumptions are highly likely to be true. | See Step 6 below. |
| 7. Compare Analytic Designs | See simulation results reported in Step 7 below. | |

[§]The revised definition of MACE using all-cause death instead of death from cardiovascular causes was chosen as the primary outcome because cause of death is not available in the RWD.
[§§]The causal gap is the difference between the true value of the causal estimand that answers the causal question and the true value of the statistical estimand that we will estimate[25].

**Figure 1: Diagram of Study Designs 1-3**

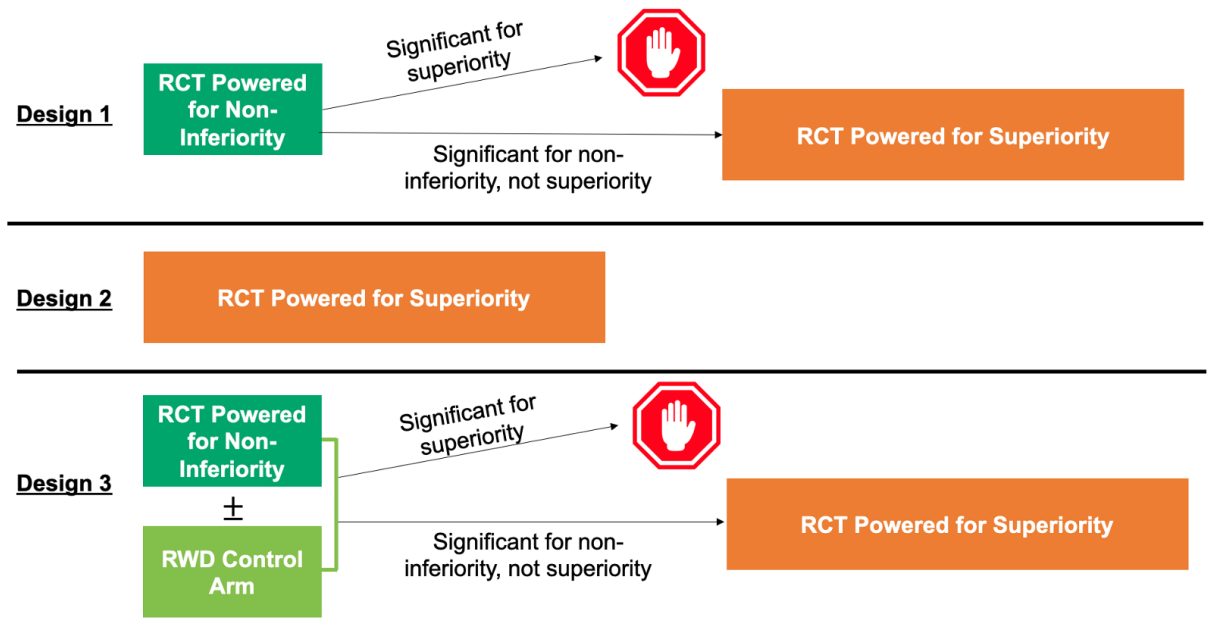

**Step 1a: Define the causal question and estimand**

The question for all study designs was: what would the difference in risk of MACE (defined as death from any cause, nonfatal MI, or nonfatal stroke) within one year be if all patients in a population consistent with the PIONEER 6 inclusion/exclusion criteria and timeframe[10], and with similar healthcare engagement, were prescribed oral semaglutide plus standard-of-care compared to if all patients were prescribed standard-of-care alone, and if censoring had been prevented for all patients? The outcome for this case study includes all-cause death (rather than cardiovascular death as in PIONEER 6) because the real-world data do not include cause of death. See Appendix 1 for the causal estimand.

## Step 1b: Specify a causal model

Next, we specify a causal model for each design describing what variables might affect treatment, censoring, or outcomes using the causal graphs[27] shown in Figure 2. For the RCTs, only the randomization procedure affects treatment assignment. As depicted in Figure 2a, health status, socioeconomic status, and related issues of healthcare access and engagement, collectively referred to as *U*, might affect both censoring and MACE. Measured pre-baseline covariates, including age, sex, race, HbA1c, high-density lipoprotein cholesterol, low-density lipoprotein cholesterol, estimated glomerular filtration rate (a marker of kidney function), prior MI, prior stroke or transient ischemic attack (TIA), prior heart failure, morbid obesity, and use of glucose-lowering medications, insulin, and CV medications, may account for some aspects of these underlying factors.

**Figure 2: Causal Graphs for Designs 1-3**

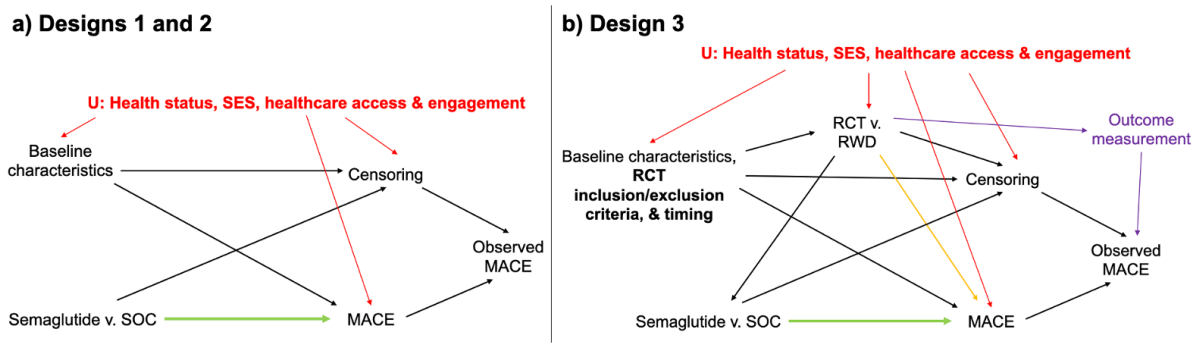

**Caption:** SOC: standard-of-care. SES: socioeconomic status. MACE: major adverse cardiovascular events. RWD: real-world data.

In the hybrid **Design 3**, participation in the RCT versus the real-world system affects treatment because RCT participation is required to receive oral semaglutide if the RWD is concurrent with the pre-approval RCT. Being in the RCT could also modify the effect of treatment or directly affect measured outcomes for reasons including closer monitoring, encouragement of adherence, variation in standard-of-care or placebo effect, or more accurate outcome measurement[28–30]. The RCT inclusion and exclusion criteria, the timeframe of RCT recruitment, health status, socioeconomic status (SES), and healthcare engagement or access may also affect trial participation.

**Step 2: Describe the observed data**

Next, we consider the data that are actually observed. **Designs 1-2** use data from one or both of the following sources: the PIONEER 6 RCT[10] and the ongoing SOUL RCT[11]. Both trials randomized participants to receive semaglutide or placebo in addition to standard-of-care. The inclusion and exclusion criteria for both trials targeted patients with T2DM and high cardiovascular risk but without unstable disease or recent use of a GLP-1RA, while PIONEER 6 also excluded recent users of pramlintide and dipeptidyl peptidase-4 inhibitors (DPP4i)[11,13]. Participants were regularly evaluated in person or by phone. Outcomes were adjudicated. Time zero was the time of randomization. The timeframe for the outcome of one year after baseline was selected because no administrative censoring occurred before that time in PIONEER 6.

Finally, the RWD considered in **Design 3** comes from Optum's Clinformatics® Data Mart (CDM), which is derived from a database of administrative health claims for members of large commercial and Medicare Advantage health plans. Clinformatics® Data Mart is statistically de-identified under the Expert Determination method consistent with HIPAA and managed according to Optum® customer data use agreements[31,32]. CDM administrative claims submitted for payment by providers and pharmacies are verified, adjudicated and de-identified prior to inclusion. This data, including patient-level enrollment information, is derived from claims submitted for all medical and pharmacy health care services with information related to health care costs and resource utilization. The population is geographically diverse, spanning all 50 of the United States.

Consistent with recommendations from the RCT DUPLICATE study, we used naive initiation of a DPP4i (defined as a new prescription following at least 90 days without a previous prescription based on AHFS codes) to enhance comparability of health care access and engagement among RWD compared to RCT controls[33]. Time zero was defined as the first time a participant met the eligibility criteria for PIONEER 6, during a calendar time window contemporaneous with PIONEER 6 recruitment, and was prescribed a DPP4i. Because oral semaglutide was approved following PIONEER 6, we only consider extra RWD control arm participants (although the method we describe can be extended to handle both external control and treatment arms).

Following the RCT Duplicate study[33], we translated and applied as many of the inclusion/exclusion criteria of PIONEER 6 as possible; details are provided in Appendix 7. This involved translating medical histories into ICD-9/ICD-10 diagnosis and procedure codes. To identify GLP1-RA and pramlintide usage, we defined continuous treatment eras as consecutive prescriptions based on AHFS codes with not more than 90-day gaps between them.

For the primary outcome, we identified nonfatal MI/stroke using ICD-9/ICD-10 diagnosis codes for inpatient visits in the first diagnosis position (see Appendix 7). All-cause death was identified from external sources as provided by Optum. We used fractures as a negative control outcome (discussed below), which we identified using ICD-9/ICD-10 diagnosis codes again for inpatient visits in the first diagnosis position.

Baseline characteristics discussed in Step 1b were determined as follows. Medical history variables were defined as described above. Medication use was identified via claims through AHFS codes, where treatment at baseline corresponded to at least one prescription in the 180 days preceding time zero. Laboratory measurements were identified through LOINC codes, where baseline values were identified as the most recent measurement prior to time zero. If the latest measurement was more than 180 days prior to time zero, then that measurement was deemed missing.

**Steps 3-4: Assess identifiability and specify a statistical estimand**

We now aim to translate the causal effect of interest (Step 1a) into a statistical estimand – a function of the observed data that we will estimate – based on our knowledge about the processes that generated our data (Step 1b). When using RCT data alone (**Designs 1 and 2**), we assume no unmeasured common causes of treatment or censoring and MACE, and adequate data support (positivity)[34,35]. These assumptions are highly likely to hold by design; while it is possible that there are unmeasured common causes of censoring and MACE, because censoring was negligible (0.3% in PIONEER 6), this would be unlikely to impact the results. A design (not considered here) in which we committed to augmenting the RCT data with external control data would require additional assumptions: no unmeasured common causes of selection into the RCT versus the RWD cohort or censoring, and MACE (no U in Figure 2b), no effect of RWD vs. RCT participation on MACE other than through effects on treatment assignment (no direct arrow from RCT vs RWD to MACE or outcome measurement in Figure 2b), and adequate data support (positivity[34,35]).

If these assumptions are not close to being satisfied, **Design 3** is likely to reject the RWD controls. The following actions were taken to improve the plausibility of these assumptions and therefore the likelihood that RWD controls would be integrated in the hybrid design: 1) selecting RWD controls with a similar disease stage and healthcare access and engagement compared to the RCT controls based on those who were prescribed an active comparator medication (DPP4i) and had relevant baseline labs and medical history recorded, 2) restricting RWD controls to the time period of PIONEER 6 recruitment to make standard-of-care more similar, and 3) selecting RWD controls whose baseline characteristics are shared by at least some RCT participants. Note that action 1 restricts the target population for the hybrid analysis to patients with relatively strong access to and engagement with the healthcare system, similar to the types of patients who are able to enroll in an RCT. Note also that use of DPP4i as an active comparator – especially considering that DPP4i use was an exclusion criterion in PIONEER 6 – also requires the assumption, supported by data, that DPP4 inhibitors do not influence cardiovascular outcomes[33]. Action 3 is necessary to avoid a violation of the positivity assumption, but doing so restricts the target population further, preventing generalization beyond the types of patients that were represented in the RCTs.

As always, when observational data are considered, it remains unlikely that the causal identification assumptions are exactly true. Yet with the above considerations, it is now plausible that the bias from integrating RWD will be small. Importantly, however, rather than relying on these assumptions holding, our proposed **Design 3** uses an estimator (described in Step 5) in which pre-specified statistical criteria are used to estimate the "causal gap", or difference between the wished-for causal effect and the statistical estimand[25] that may be estimated by pooling RCT and RWD[17]. The estimator only selects the RWD for inclusion when the combined impact of any deviations from these assumptions is unlikely to impact accurate inference[17]. This approach greatly mitigates, but does not eliminate, the threat of misleading inference, which is further evaluated in Steps 6 and 7. Appendix 1 has mathematical expressions for the statistical estimands that are equivalent to the wished-for causal effects (causal estimands) under the identification assumptions, which are discussed in detail in Appendix 2.

**Step 5: Choose a statistical model and estimator**

For all designs, we use a statistical model that avoids any assumptions not firmly grounded in design knowledge (e.g., treatment randomization). In **Designs 1 and 2**, because censoring was negligible, we estimated the unadjusted risk difference between arms among persons with follow-up through one year. For **Design 3**, we used the experiment-selector cross-validated targeted maximum likelihood estimator (ES-CVTMLE)[17]. Cross-validated Targeted Maximum Likelihood Estimation (cv-TMLE) is a robust, efficient approach that incorporates machine learning using internal sample splitting (cross-validation)[23,36–38]. This allows it to flexibly adjust for covariates without introducing new assumptions, improving precision and potentially reducing bias, while preserving inference[23,36–38]. The ES-CVTMLE extends this method to evaluate and integrate external RWD controls[17]. Specifically, it uses pre-specified statistical criteria to evaluate the RWD and integrate them with the RCT data only when their inclusion is unlikely to worsen confidence interval coverage.

The ES-CVTMLE estimates the bias created by augmenting the RCT with RWD in two ways: 1) by comparing the control arms (specifically the conditional mean outcomes) between the two settings and 2) through the use of a negative control outcome (NCO)[17]. From the available options, we chose fractures as an NCO because it is generally serious enough to require medical attention for those with access, is associated with SES[39], and is recorded in a manner similar to the primary outcome. Studies prospectively designed to include an NCO could consider alternate choices.

In this case study, we chose the ES-CVTMLE as the estimator for the hybrid design because it relies on few statistical assumptions, incorporates an estimate of bias based on an NCO, and adjusts confidence interval widths based on the estimated magnitude of bias. While the focus of this case study is on using simulations to compare study designs, similar simulations could also be used to compare different potential estimators. Appendices 3-6 provide further details about estimation.

**Step 6: Specify a procedure for sensitivity analysis**

For the RCT-only analyses (**Designs 1 and 2**), because identification assumptions are likely to hold and we avoid any unsupported statistical assumptions, there are few threats to inference, and sensitivity analyses may be unnecessary. In **Design 3,** for the hybrid RCT-RWD analysis, the ES-CVTMLE is designed to conservatively protect inference without additional assumptions. However, as with all estimators that aim to estimate bias from including external data in order to decide whether or how to include RWD in a hybrid analysis[18–22,40], it still runs a small but real risk of failing to reject inappropriate external controls. To evaluate the extent to which this threatens inference, a tipping point analysis can be carried out to evaluate how much of a causal gap for the combined RCT-RWD study is needed to make the confidence interval from Step 5 non-significant[41–44]. Additional simulation-based approaches, incorporating estimates of the plausible causal gap, can also be used to augment sensitivity analyses.

**Step 7: Compare analytic designs using simulations**

We compare **Designs 1-3** using simulations that closely mimic our true study designs (Appendix 4). For **Designs 1 and 3**, we simulate data from a small RCT aiming to mimic PIONEER 6 (RCT1). For **Design 1**, we use an unadjusted estimate of the difference in risk between arms of RCT1. For **Design 3**, we consider not only the simulated RCT1 but also a simulated "real-world" dataset aiming to mimic CDM, and we use the ES-CVTMLE to estimate the risk difference. In both **Designs 1 and 3**, if the null hypothesis is rejected at this first stage, then we use this initial estimate as our final effect estimate. If the null hypothesis is not rejected, then we simulate data from a larger trial aiming to mimic a superiority trial (RCT2) and estimate the risk difference using an unadjusted estimate. In **Design 2,** we simply simulate RCT2 and use the unadjusted effect estimate.

Because we consider non-randomized data in **Design 3**, it is possible that the confidence interval coverage will be lower than in **Designs 1-2** if the causal identification assumptions from Step 3 are violated. Yet it may be acceptable if, in some contexts (i.e., bias of a certain magnitude), coverage is below 95% if there is sufficient benefit to patients of **Design 3** over **Designs 1-2**. While there are many potential ways to quantify benefit to patients, for this case study, we estimate the number of patient-years during which participants are precluded from starting a GLP1-RA (by being in an RCT control arm), averaged over 1000 iterations of this simulation. We also report power to reject the null hypothesis using $\alpha=0.05$.

We evaluate **Design 3** when the magnitude of bias introduced by including the simulated RWD in the analysis is zero and when it is one of ten potential magnitudes in either the positive or negative direction, ranging up to $\pm 2.1\%$. Appendix 4 describes the rationale for this maximum bias. In this primary simulation, the effect of unmeasured factors causing bias is the same on the relationship between semaglutide and MACE as it is on the relationship between semaglutide and the NCO. Appendix 5 shows the results of the same simulation both when the NCO is not considered (leading to more conservative inference) and when the NCO is considered but the unmeasured factors causing bias for the relationship between semaglutide and MACE have no effect on the NCO, mimicking a worst-case scenario for violations of both the causal identification assumptions and the assumptions needed to measure bias using an NCO.

**Results**

**Simulation Results**

Figure 3 shows the results of 1000 iterations of the simulation comparing **Designs 1-3**. **Design 1** and **Design 2** had similar characteristics. Simulated **Design 1** had 95% CI coverage of 0.949, power of 0.842, and an average of 4,765 patient-years during which a GLP1-RA was precluded. Simulated **Design 2** had coverage of 0.955, power of 0.764, and an average of 4,750 patient-years during which a GLP1-RA was precluded.

**Figure 3: Simulation Results by Study Design with Different Amounts of RWD Bias**

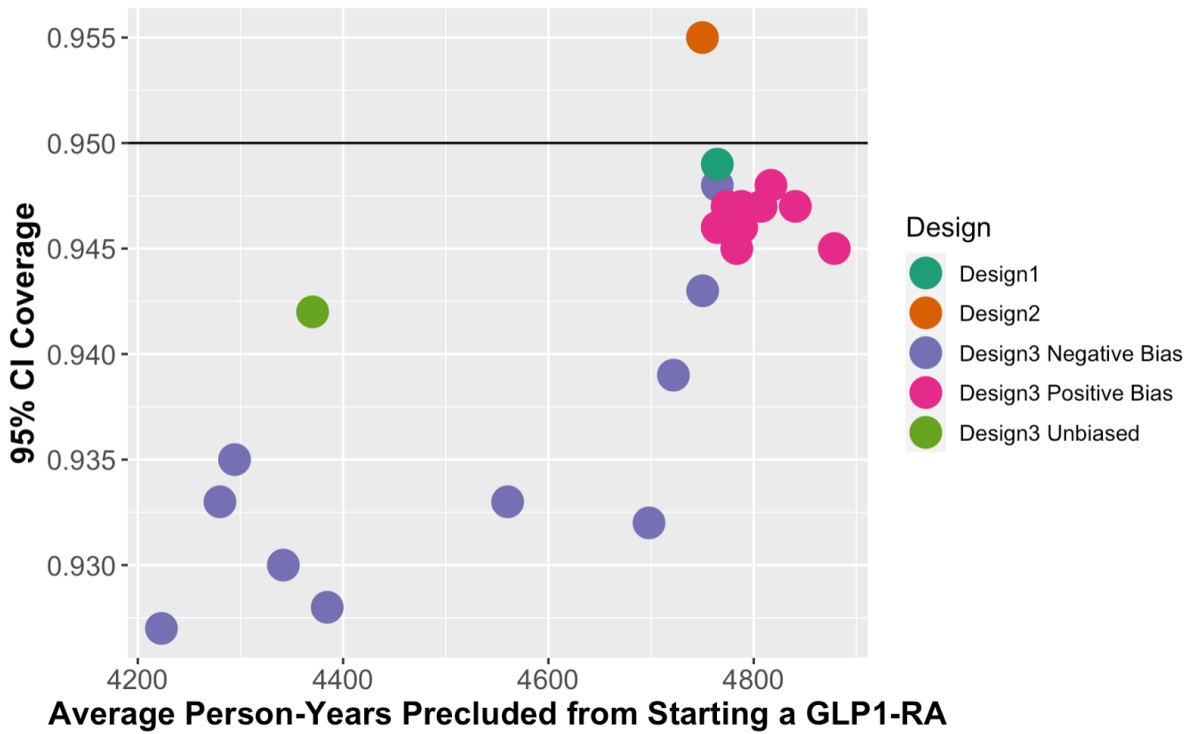

**Caption:** Purple represents 10 simulated magnitudes of bias away from the null. Pink represents 10 simulated magnitudes of bias towards the null. CI: Confidence Interval.

The tradeoffs between **Design 3** and **Designs 1 and 2** depended on the direction of bias introduced by the RWD. With unbiased simulated RWD, **Design 3** had coverage of 0.942, had power of 0.866, and resulted in an average of 394 fewer participant-years during which patients were precluded from starting a GLP1-RA compared to **Design 1**. In other words, on average, 8.3% fewer people would have spent one year during which their doctor avoided prescribing a GLP1-RA if **Design 3** were chosen and unbiased RWD were available compared to if **Design 1** were chosen.

Positive bias (towards the null) represents scenarios in which the introduction of RWD lowers the estimated risk of MACE among control arm participants. This could happen if MACE was not well-recorded in the RWD. Simulated positive bias led to coverage ranging from 0.945 to 0.948, power ranging from 0.842 to 0.847, and an average of 0 to 114 extra participant-years during which prescription of a GLP1-RA was discouraged compared to **Design 1**, with the largest increases for intermediate magnitudes of bias. The increase in person-years without GLP1-RA access occurred because RWD with bias towards the null were included in a small number of simulation iterations, triggering a second RCT. A study comparing outcomes recorded in CDM to outcomes recorded following an RCT protocol could be conducted to assess the likelihood that the proposed hybrid design would truly result in this spectrum of positive RWD bias.

Negative bias (away from the null) represents scenarios in which the introduction of RWD raises the estimated risk of MACE among control arm participants. This could happen if participants in the real world had worse health outcomes than trial participants due to differences described in Section 2 and these differences were not adequately detected through comparison of RWD and RCT control arms or via negative control. Simulated negative bias led to coverage ranging from 0.927 to 0.948, power ranging from 0.847 to 0.873, and an average of 0 to 542 fewer participant-years during which prescription of a GLP1-RA was discouraged compared to **Design 1.** For comparison, if one had used a naive estimator that simply assumed that external control data were appropriate to integrate (i.e., the stringent assumptions discussed in Step 3 held), and simply pooled the RCT and most biased real-world data, coverage would have been 0.126. The ES-CVTMLE thus provided significant (though imperfect) protection against integration of biased RWD in this simulation.

The possibility of bias away from the null is plausible, though quantifying how much bias might be expected within the range represented by this simulation (0-2.1%) is challenging. Nonetheless, by objectively quantifying these differences between proposed designs, investigators can explicitly discuss these tradeoffs with stakeholders such as patient groups and regulatory agencies when deciding which trial design to choose.

**Real Data Analysis: The estimated effect of oral semaglutide on MACE from PIONEER 6, considering augmentation with additional CDM RWD controls**

The actual results of **Designs 1 and 2** await completion of the SOUL trial. Below, we carry out **Design 3** using data from PIONEER 6 and the CDM external control arm described in Step 2 above. We also report the results of an unadjusted estimator for the difference in the risk of MACE among PIONEER 6 active and control arm participants.

After applying the inclusion and exclusion criteria described in Step 2 and depicted in Figure 4, the CDM cohort consisted of 2483 participants. Table 2 lists baseline demographics, medical history, medication use, outcome missingness, and MACE and NCO event rates for the PIONEER 6 semaglutide and placebo arms, as well as for the CDM external control arm. The outcome was missing for 0.3% of PIONEER 6 participants and 16% of CDM participants. The negative control outcome was missing for 0.3% of PIONEER6 participants and 17% of CDM participants. Compared to the PIONEER 6 control arm, the CDM controls were slightly older (more participants in the 70-80 year-old range), had a higher percentage of females, had a lower proportion of previous MI or stroke but a higher proportion of previous heart failure, and had a different distribution of baseline medication use.

Figure 4: Selection of CDM External Control Group

4a) Flow Diagram
Full database, at least 180 days of observations, T2DM

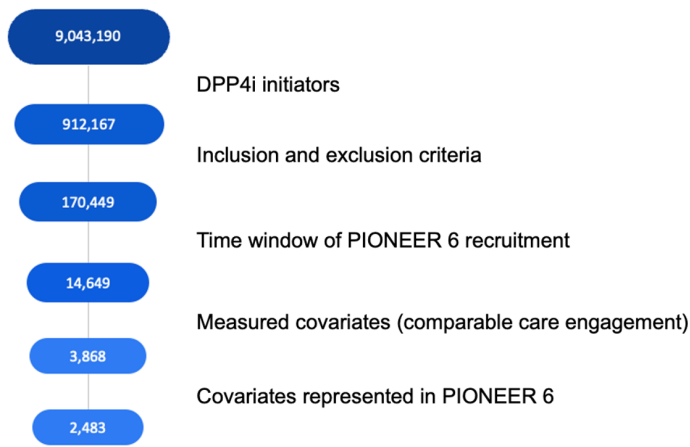

4b) Timing of RCT Randomization and CDM Active Comparator Initiation

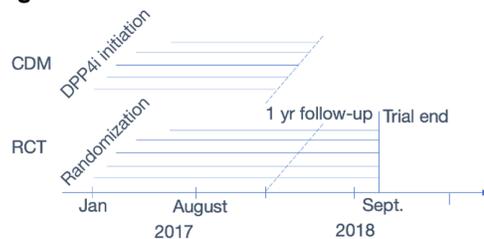

**Caption:** T2DM: Type 2 Diabetes Mellitus. DPP4i: Dipeptidyl Peptidase 4 inhibitor.

**Table 2: Baseline Characteristics, Outcome Missingness, and Event Rates for PIONEER 6 and CDM**

| | CDM RWD control arm (n=2483) | PIONEER 6 placebo arm (n=1564) | PIONEER 6 semaglutide arm (n=1574) |
|---|---|---|---|
| MACE rate - % | 4.5 | 4.2 | 2.9 |
| NCO Rate - % | 0.73 | 0.77 | 0.45 |
| Age - years, mean (SD) | 69.2 (6.3) | 66.4 (7.1) | 65.9 (7.1) |
| Female sex - % | 42.7 | 31.2 | 31.9 |
| Race | | | |
|     White - % | 44 | 72 | 72 |
|     Black - % | 11 | 7 | 6 |
|     Other - % | 45 | 21 | 22 |
| HbA1c - %, mean (SD) | 8.0 (1.4) | 8.2 (1.6) | 8.2 (1.6) |
| LDL cholesterol - mg/dl, mean (SD) | 84.1 (28.3) | 84.8 (32.4) | 83.9 (34.0) |
| HDL cholesterol - mg/dl, mean (SD) | 44.1 (9.6) | 41.6 (10.7) | 41.9 (11.0) |
| eGFR - ml/min/1.73 m2, mean (SD) | 74.3 (19.4) | 74.2 (20.9) | 74.2 (21.1) |
| Previous MI - % | 13.9 | 36.9 | 35.3 |
| Previous stroke/TIA - % | 11.8 | 16.6 | 15.2 |
| Previous heart failure - % | 20.4 | 12.4 | 11.9 |
| Morbid obesity - % | 16.4 | 12.3 | 12.2 |
| Glucose-lowering medication (metformin, SU, TZD, SGLT2i) | 73.0 | 83.9 | 83.9 |
| Insulin | 14.9 | 61.2 | 61.2 |
| Cardiovascular medication (antihypertensives, lipid-lowering, anti-thrombosis, diuretics) | 91.5 | 98.9 | 98.9 |
| Outcome missingness - % | 15.8 | 0.3 | 0.3 |

| NCO missingness -% | 17.2 | 0.3 | 0.3 |

**Caption:** LDL: low-density lipoprotein, HDL: high-density lipoprotein, eGFR: estimated glomerular filtration rate, MI: myocardial infarction, TIA: transient ischemic attack, SU: sulfonylurea, TZD: thiazolidinedione, SGLT2i: sodium/glucose cotransporter-2 inhibitor

As shown in Figure 5, the estimated difference in the risk of MACE by 1 year based on the unadjusted estimator conducted using PIONEER 6 data alone was -1.30%-points (95% CI -2.60 to 0.00%-points). This result is closer to statistical significance than the primary result reported for the PIONEER 6 trial (hazard ratio 0.79; 95% CI 0.57 to 1.11 [10]) because the primary analysis a) evaluated the hazard ratio including all timepoints instead of the risk difference by one year and b) evaluated a composite outcome that included death from cardiovascular causes instead of death from all causes. Nonetheless, the confidence interval for the result of this modified analysis still includes zero.

**Figure 5: Estimated Difference in 1-Year Risk of MACE for PIONEER 6 and Hybrid Design**

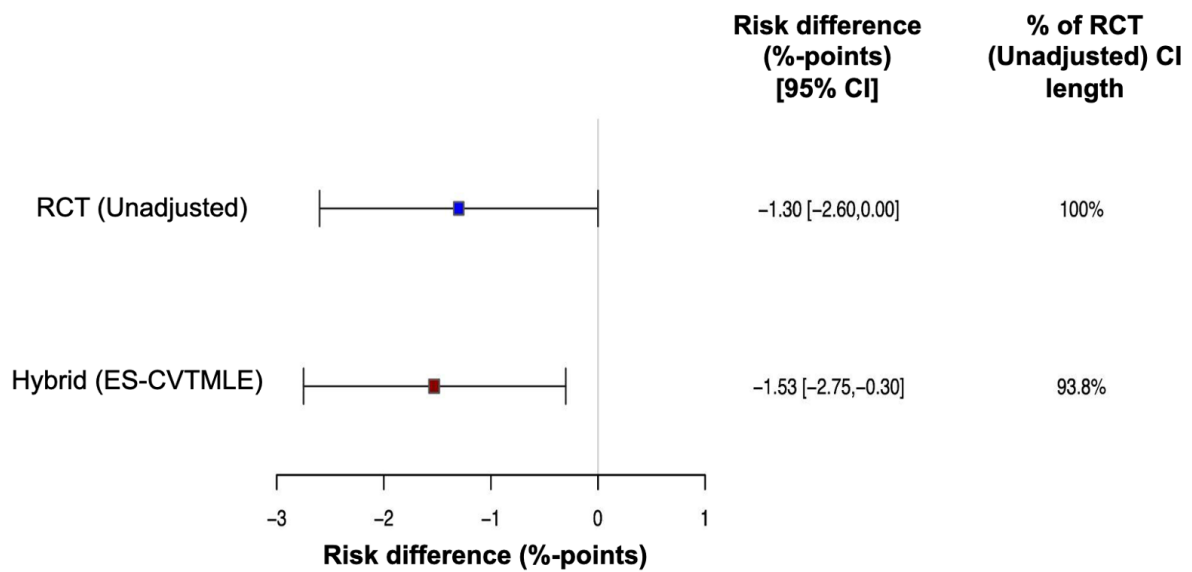

**Caption:** CI: Confidence Interval. ES-CVTMLE: Experiment-Selector Cross-Validated Targeted Maximum Likelihood Estimator.

Hybrid **Design 3** resulted in an estimated risk difference of -1.53%-points (95% CI -2.75 to -0.30%-points), providing evidence in support of the superiority of oral semaglutide versus standard-of-care for the prevention of MACE. The two primary differences between the risk difference estimates from PIONEER 6 alone compared to the hybrid analysis are narrower confidence intervals and a small negative shift in the point estimate. Narrower confidence intervals are expected as the CDM RWD were included in the analysis in 84% of internal sample splits, leading to increased efficiency.

The small shift in the point estimate could occur for three main reasons. First, the magnitude of the shift is well within what might be expected by chance alone. Second, the shift may be due to subtle changes in the target population that arise from including external controls (even those with equivalent eligibility criteria (Table 2)). Finally, as discussed above and illustrated using simulations, the potential for some residual bias remains.

Simulations, such as those presented in Step 7, can help stakeholders to weigh the tradeoffs between the benefits and downsides to patients of the different study designs against the risk of misleading inference. Had this design been proposed prior to running SOUL, explicit evaluation of these tradeoffs could have facilitated discussions of whether the evidence produced by the proposed hybrid design would be sufficient to inform decisions regarding label extensions for oral semaglutide for the secondary indication of cardiovascular risk reduction.

**Discussion**

In this case study of the effect of oral semaglutide on major adverse cardiovascular events, we demonstrate an application of the *Causal Roadmap* to a hybrid RCT-RWD trial that considers integration of data from multiple sources. We also implement the extension proposed in the companion article[26] to use the *Causal Roadmap* to compare different potential study designs using simulations. Both the FDA guidance on complex innovative trial designs[45] and the FDA guidance on adaptive designs[46] suggest the utility of simulations for comparing alternative design choices. The simulated results demonstrate that selection of one potential design over another may depend on the direction of the bias introduced by RWD and suggest the importance of conducting studies to evaluate the potential for bias with integration of different RCT and RWD sources.

Our estimate of the difference in the risk of MACE by one year with oral semaglutide versus standard-of-care attained statistical significance, providing evidence in support of the superiority of oral semaglutide as compared to standard-of-care alone. Regulatory decisions regarding whether to extend the label of oral semaglutide to include the secondary indication of cardiovascular risk reduction will await the results of the SOUL trial. Nonetheless, hybrid RCT-RWD studies could potentially be used in the future to provide additional relevant information to regulatory agencies for secondary indications for a wide variety of disease processes – in this case, a common adult disease rather than a pediatric or rare disease as have been highlighted in previous applications of hybrid trial methodologies[15,16].

While this study aimed to quantify tradeoffs between three proposed designs, other approaches could be considered. Hybrid RCT-RWD designs may adapt the probability of randomization to active treatment based on the efficiency gains that are achieved by integrating RWD and also on the probability of superiority compared to placebo[19,47,48], potentially leading to even less patient-time on an inferior product. Power could also have been higher if oral semaglutide had been available in the CDM dataset for the specified time period – resulting in integration of both extra treatment and control arm participants – or if more RWD controls were available.

A key limitation of this study is that it was planned after PIONEER 6, and thus the simulations and analysis plan were not pre-specified without knowledge of trial results. In RWD studies aiming to support policy and/or regulatory decision-making, all design and analysis decisions should be pre-specified before effects are estimated from any of the data sources that are considered[26]. The *Causal Roadmap* supports a rigorous design process and reporting structure to ensure pre-specification of components needed to support the validity of causal inferences drawn from such designs. The current case study provides a detailed worked example of this process.

This study reports one of many potential metrics aimed at quantifying the benefits and drawbacks of different study designs from the perspective of patients. While recommendations have been proposed regarding the elicitation of patient perspectives to inform medical product development[49,50], further guidance on the most relevant metrics of patient benefit as well as best practices for collaboratively weighing tradeoffs between different metrics of design performance is warranted. Decisions regarding these tradeoffs are context-specific[45], but following the *Causal Roadmap* may help investigators to collate and quantify the information that is most relevant for discussions with stakeholders. For this case study of semaglutide and cardiovascular outcomes, application of the *Causal Roadmap* prompts us to ask whether more patients could have benefited from receiving a GLP1-RA sooner if a hybrid RCT-RWD approach had been taken.

**Acknowledgements:** We would like to thank Dr. Richard Pratley, Dr. Nicky Best, Dr. Josh Fessel, Dr. Alex D'Amour, Dr. Charlie Barr, and Dr. Rima Izem for their comments on this case study during the Forum on the Integration of Observational and Randomized Data (FIORD) meeting working group sessions. We would like to thank Dr. Hana Lee for her comments but note that her contribution was technical only and not related to the study currently under an Investigational New Drug application. We would also like to thank the sponsors of the FIORD workshop, including the Forum for Collaborative Research and the Center for Targeted Machine Learning and Causal Inference (both at the School of Public Health at the University of California, Berkeley), and the Joint Initiative for Causal Inference. This research was funded by a philanthropic gift from the Novo Nordisk corporation to the University of California, Berkeley to support the Joint Initiative for Causal Inference.

**Disclosures: LED** reports tuition and stipend support from a philanthropic gift from the Novo Nordisk corporation to the University of California, Berkeley to support the Joint Initiative for Causal Inference. **EF, JMT, KKBC, HR, and KK** are full-time employees of Novo Nordisk A/S and own stocks in Novo Nordisk A/S. **KKBC** is also affiliated with Steno Diabetes Center Copenhagen, Denmark. **JBB** reports contracted fees and travel support for contracted activities for consulting work paid to the University of North Carolina by Novo Nordisk; grant support by Dexcom, NovaTarg, Novo Nordisk, Sanofi, Tolerion and vTv Therapeutics; personal compensation for consultation from Alkahest, Altimmune, Anji, AstraZeneca, Bayer, Biomea Fusion Inc, Boehringer-Ingelheim, CeQur, Cirius Therapeutics Inc, Corcept Therapeutics, Eli Lilly, Fortress Biotech, GentiBio, Glycadia, Glyscend, Janssen, MannKind, Mellitus Health, Moderna, Pendulum Therapeutics, Praetego, Sanofi, Stability Health, Terns Inc, Valo and Zealand Pharma; stock/options in Glyscend, Mellitus Health, Pendulum Therapeutics, PhaseBio, Praetego, and Stability Health; and board membership of the Association of Clinical and Translational Science. **MvdL** reports that he is a co-founder of the statistical software start-up company TLrevolution, Inc. **MvdL and MP** report personal compensation for consultation from Novo Nordisk.

**Supplementary Appendices**

**Appendix 1: Mathematical Notation for Causal and Statistical Estimands**

As described in Table 1, the causal question of interest is: what would the difference in risk of MACE (defined as death from any cause, nonfatal MI, or nonfatal stroke) within one year be if all patients in a population consistent with the PIONEER 6 inclusion/exclusion criteria and timeframe[1], and with similar healthcare engagement, were prescribed oral semaglutide plus standard-of-care compared to if all patients were prescribed standard-of-care alone, and if censoring had been prevented for all patients?

The two intervention variables that are modified in our treatment strategies are $A$ – an indicator of prescribing patients oral semaglutide in addition to standard-of-care ($A=1$) or standard-of-care ($A=0$) – and $C$ – an indicator of whether the participants were censored before one year. We denote our outcome of MACE by 1-year as $Y$. Because some participants are censored, the observed outcome, $Y^\star$, is the true value of MACE for those who were not censored and whose outcomes were measured correctly and missing for those who were censored.

We then define the following potential outcomes[2,3]; $Y^{a=1,c=0}$ is the one-year MACE status an individual would have had if they had been prescribed oral semaglutide in addition to standard-of-care and not been censored, and $Y^{a=0,c=0}$ is the one-year MACE status an individual would have had if they had been prescribed standard-of-care and not been censored. The simplest mathematical representation of a causal estimand that answers our question is given by the causal risk difference:

$$E(Y^{a=1,c=0} - Y^{a=0,c=0}).$$

Note that this causal risk difference is defined with respect to a specific target population. Despite efforts to ensure comparability between the RCT population and external control RWD, our approach acknowledges that the RCT and RWD populations may nonetheless have different distributions of baseline characteristics. Because the proposed estimator (ES-CVTMLE) only augments the control arm when the RWD meet pre-specified criteria (evaluated across multiple internal sample splits), the exact target population to which the causal risk difference applies will depend on the extent to which these criteria are met and the RCT standard-of-care arm is augmented with RWD.

More formally, let *S* be a variable describing study participation, where *S=0* indicates that an individual participated in an RCT and *S=1* indicates that an individual participated in the real-world healthcare system. **Designs 1 and 2** only utilize RCT data, and so in these designs, we can only evaluate the causal risk difference within the RCT context and for a target population represented by the RCT participants. We can rewrite the causal parameter to represent the causal risk difference (not adjusted for baseline characteristics) in a way that makes explicit that it refers to the RCT context:

$$\psi^*_{RCT,unadj} = E(Y^{a=1,c=0} - Y^{a=0,c=0}|S=0).$$

With baseline covariates, W, the adjusted causal risk difference for the RCT context and target population is:

$$\psi^*_{RCT,adj} = E_{W|S=0}[E(Y^{a=1,c=0} - Y^{a=0,c=0}|W, S=0)].$$

We note that the true causal risk difference in the non-inferiority and superiority trials could be different if they had different inclusion and exclusion criteria or if there were changes over time in the background standard-of-care. For simplicity, however, we will consider that the non-inferiority and superiority RCTs target the same causal parameter, $\psi^*_{RCT,unadj}$.

In hybrid **Design 3,** we consider integrating extra RWD controls with our non-inferiority trial and only run the superiority RCT if the null hypothesis of the superiority RCT is not rejected in the hybrid analysis. The ES-CVTMLE adjusts for baseline covariates whether the RWD is included or rejected, so if the hybrid design rejects the RWD, analyzing the non-inferiority RCT only, then the causal target parameter is $\psi^*_{RCT,adj}$.

In contrast, if the hybrid design does select to augment the non-inferiority trial with extra RWD controls, this may modify the target population if the RWD controls have a different (though overlapping) distribution of baseline covariates compared to the RCT population. Inclusion of RWD controls may also modify the target parameter if the true effect of oral semaglutide versus standard-of-care is different in the RCT and RWD contexts. The causal risk difference in the combined RCT plus RWD experiment that integrates *S=0* and *S=1* is given by:

$$\psi^*_{RCT,RWD} = E_{W|S\in\{0,1\}}[E(Y^{a=1,c=0} - Y^{a=0,c=0}|W, S \in \{0,1\})].$$

In the hybrid design, we use the data from the real-world source and the non-inferiority trial to decide whether to estimate the causal risk difference for the RCT context and target population ($\psi^*_{RCT,adj}$) or the causal risk difference for the hybrid RCT-RWD context and target population ($\psi^*_{RCT,RWD}$), where either parameter represents an answer to our question for that particular population and context.

We will ultimately use the experiment-selector CV-TMLE[4] to analyze the results of the hybrid trial. This method uses cross-validation to separate the part of the data that is used to choose whether to attempt to estimate $\psi^*_{RCT,adj}$ or $\psi^*_{RCT,RWD}$ and the part of the data that is used for estimation of the corresponding risk difference. The decision of whether to augment the RCT with external control data may differ in different cross-validation folds. The causal estimand would then be interpreted as the causal risk difference for a target population that is a weighted average of the RCT population and external control RWD population. More formally, let the target parameter chosen for a given fold, *v*, be $\psi^*_v$. The overall causal target parameter for the hybrid design is then the average of the causal target parameters selected in each fold. For example, with ten cross-validation folds, the causal target parameter would be

$$\psi^*_{hybrid} = \frac{1}{10} \sum_{v=1}^{10} \psi^*_{v.}$$

Please see Dang et al. (2022)[4] for further details of this methodology.

Using the g-formula[5], we may define the statistical estimands (functions of the observed data) that are as close as possible to the causal effects of interest for each study design, where

$$\psi_{RCT,unadj} = E(Y^*|C = 0, A = 1, S = 0) - E(Y^*|C = 0, A = 0, S = 0),$$

$$\psi_{RCT,adj} = E_{W|S=0}[E(Y^*|C = 0, A = 1, W, S = 0) - E(Y^*|C = 0, A = 0, W, S = 0)],$$

$$\psi_{RCT,RWD} = E_{W|S\in\{0,1\}}[E(Y^*|C = 0, A = 1, W, S \in \{0,1\}) \\ - E(Y^*|C = 0, A = 0, W, S \in \{0,1\})],$$

$\psi_v$ is whichever of $\psi_{RCT,adj}$ or $\psi_{RCT,RWD}$ was selected in cross-validation fold $v$, and

$$\psi_{hybrid} = \frac{1}{10}\sum_{v=1}^{10}\psi_v.$$

**Appendix 2: Assessment of Plausibility of Causal Identification Assumptions**

First, we consider whether the causal effect is identified in **Designs 1 and 2** based on the directed acyclic graph (DAG) in Figure 2a, using the backdoor criterion[6]. Because treatment was randomized, there are no unmeasured common causes of treatment and the outcome. Furthermore, because censoring is minimal in the RCTs (it was truly 0.3% by one year in PIONEER 6)[1], the magnitude of bias that could result from potential unmeasured common causes of censoring and MACE is likely to be negligible. For these reasons, we expect to identify the causal effect of interest using **Designs 1 and 2**.

As discussed in the main text, **Design 3** does not assume that the causal effect of interest is identified in the pooled RCT and RWD. However, taking steps to improve the plausibility of causal identification assumptions for the combined data also increases the likelihood that RWD will be integrated in the hybrid design. Again using the backdoor criterion[6] and our DAG in Figure 2b, we consider possible reasons for a causal gap in an analysis of the pooled RCT and RWD. First, there would be a causal gap if being in the RCT versus the real world affected outcomes outside of the effect due to prescribing either oral semaglutide or standard-of-care; this is certainly possible for the reasons described in Step 2 above.

If we were able to conduct a pragmatic clinical trial for the randomized component of our hybrid design in which the trial aimed to mimic real-world care as closely as possible outside of baseline treatment randomization, then it would be more likely that trial participation only affected outcomes through treatment assignment[7,8]. In this case, however, we are not able to consider a pragmatic RCT because the current FDA draft guidance on "Evaluating the Safety of New Drugs for Improving Glycemic Control"[9] requires a sufficient number of phase 3 clinical trial patient-years on the medication of interest during which time CV outcomes are evaluated by adjudication to evaluate cardiovascular safety. Instead, we may attempt to select RWD controls who at least have similar healthcare access and engagement compared to RCT participants (discussed below).

There would also be a causal gap for the pooled RCT and RWD analysis if there were unmeasured common causes of trial participation and MACE or of censoring and MACE, and we expect a larger amount of censoring in the real world compared to the RCT. We apply the same inclusion and exclusion criteria and timeframe for the RWD and RCT controls, yet the question remains whether the measured baseline characteristics that are indicative of demographics, baseline health status, and treatment are sufficient to adjust for common causes of our intervention variables and outcome.

To try to minimize the amount of bias that would be introduced by integrating RCT and RWD controls, we consider a further restriction of the CDM cohort to select patients who are likely to be at a similar disease stage with similar healthcare access and engagement compared to the RCT participants. Selecting RWD patients prescribed DPP4is is one method of making disease stage and engagement comparable[10]. We also exclude CDM patients with missingness in the baseline covariates, expecting that patients for whom this laboratory and medical history data is not recorded might not be followed as closely by their providers as patients in the RCT.

Additionally, there would be a causal gap if assignment to standard-of-care in the trial and the RWD were not equivalent in terms of their effects on MACE. The most obvious reason this might not be true is that participants in the RCT were prescribed an inactive placebo pill and were not prescribed a DPP4i based on exclusion criteria, while the RWD participants were prescribed a DPP4i as an active comparator. The question then is whether the effect of being assigned placebo is different from the effect of being prescribed a DPP4i on the outcome of MACE. In the RCT Duplicate study, DPP4is were chosen as a "proxy for placebo" relative to the outcome of MACE in studies of GLP-1RAs "because they are antidiabetic treatments that have similar indications to the treatments under study, but they are not known to have any effect on the cardiovascular outcomes of interest based on recent evidence"[10]. If this reasoning is correct, assignment to placebo should have the same effect as assignment to a DPP4i for the primary outcome. Another question is whether the background standard-of-care that patients receive is equivalent in the RCT and the RWD. While it is possible that there are differences between the standard-of-care provided by trial versus non-trial clinicians, we attempted to ensure that "standard-of-care" would be as similar as possible by restricting the CDM cohort to the same time period as PIONEER 6.

Finally, to identify a causal effect in the combined RCT and RWD, we need sufficient data support. In other words, participants in any stratum of measured confounders must have a positive probability of being assigned to either intervention strategy: oral semaglutide and not being censored or standard-of-care and not being censored. This assumption is also known as the positivity assumption[11,12]. Because we only add extra RWD controls, including any RWD participant whose particular combination of measured potential confounding variables was not shared by RCT participants would violate the positivity assumption. We solve this problem by limiting the CDM cohort to participants whose baseline covariates were within the range of baseline covariates represented in the trial population.

**Appendix 3: Estimation of the Causal Gap**

The first estimate of the causal gap used by the ES-CVTMLE compares conditional mean outcomes between RCT and combined RCT-RWD controls. The statistical estimand for this causal gap parameter is given by

$$\Psi^{\#} = E_{W|S\in\{0,1\}}\big[E[Y^\star|C=0, A=0, S=0, W] - E[Y^\star|C=0, A=0, S\in\{0,1\}, W]\big].$$

The ES-CVTMLE estimates $\Psi^{\#}$ using targeted maximum likelihood estimation[13,14], but the precision of the estimate depends on the sample size of the RCT. In a given sample dataset, the estimate of $\Psi^{\#}$ will not be exactly equal to the true causal gap because of finite sample variability. Nonetheless, $\Psi^{\#}$ represents our best estimate of the causal bias that would be introduced by including RWD controls in the analysis. See Dang et al. (2022)[4] for more details.

We also estimate the causal gap as the estimated average treatment effect on a negative control outcome (NCO). NCOs are not affected by the treatment but ideally should be affected by as many of the factors that lead to violations of identification assumptions as possible[15]. Any estimated association between treatment and the NCO is thus due either to a causal gap or due to finite sample variability.

**Appendix 4: Simulation set-up**

The data for the simulation were generated as follows. First, we generate data to mimic a non-inferiority RCT (RCT1) of sample size $n_1$=3183, twenty-one different "real-world" datasets (RWD) of sample size $n_2$=2483, and a superiority RCT (RCT2) of sample size $n_3$=9500. In the two "RCT" datasets, treatment is randomized with probability 0.5. In the "RWD", all participants receive *A=0*. Two baseline covariates, *W1* and *W2* are drawn from $Normal(\mu = 0, \sigma = 1)$ distributions for participants from all studies.

We generate 21 potential levels of bias in the "RWD" as follows. *B* is a variable that introduces bias when non-zero. The value of *B* is zero for the two "RCT" datasets and the unbiased "RWD" dataset. For the remaining 20 "RWD" datasets, the value of *B* ranges from positive $\frac{1}{10} * 0.65$ to $\frac{10}{10} * 0.65$ in increments of $\frac{1}{10} * 0.65$ and from $\frac{-1}{10} * 1.7$ to $\frac{-10}{10} * 1.7$ in increments of $\frac{-1}{10} * 1.7$. These values were chosen because, due to the properties of the $logit^{-1}$ function, this range of values of *B* leads to true bias as large as ±2.1%. This maximum magnitude of bias in either direction was chosen so that the true bias minus two times the standard error of the bias estimator would be larger than the standard error of the risk difference TMLE estimator for the RCT alone. Because the ES-CVTMLE will select the combination of RCT and RWD if the estimated squared bias plus the variance of the TMLE risk difference estimator for the combined data is smaller than the estimated squared bias plus the variance of the TMLE risk difference estimator for the RCT alone, these magnitudes of bias include the full spectrum of magnitudes for which we would expect that RWD might be included in the analysis in some simulation iterations.

The primary outcome, *Y*, is generated as follows:

$$Y \sim Bernoulli(p = logit^{-1}(-3.33 + 0.2 * W1 - 0.4 * W2 + U_y + B))$$

where $U_y \sim Normal(\mu = 0, \sigma = 0.5)$. This equation was designed so that the overall probability of MACE would be similar to the true probability of MACE in the PIONEER 6 placebo arm (4.2%). Adding extra random error, $U_y$, means that the baseline covariates are not very predictive of the outcome, which is common with relatively rare binary outcomes measured years after baseline.

In order that the magnitude of the effect of B on the relationship between the treatment and the negative control outcome (NCO) be similar to the magnitude of the effect of B on the relationship between the treatment and the true outcome, but to make sure that the primary and negative control outcomes are not too tightly correlated, we let

$$NCO \sim Bernoulli(p = logit^{-1}(-3.33 + 0.2 * W1 - 0.4 * W2 + U_{nco} + B))$$

where $U_{nco} \sim Normal(\mu = 0, \sigma = 0.5)$ but is independent of $U_y$. The simulation in Appendix 5 describes an alternate, "worst-case" simulation in which B has no effect on the NCO.

We also generate some missing outcomes, where the indicator that the outcome is censored ($C = 1$) and the indicator that the NCO is censored ($C_{nco} = 1$) are generated as follows:

$$C \sim Bernoulli(p = \left(1 - logit^{-1}(2.2 + W1 - W2 + 4.5 * I(S = 0))\right))$$

$$C_{nco} \sim Bernoulli(p = \left(1 - logit^{-1}(2.2 + W1 - W2 + 4.5 * I(S = 0))\right))$$

where *S=0* indicates one of the simulated RCT datasets. These equations for outcome missingness were designed to approximate the true rates of outcome missingness in the RCT context (0.3% for PIONEER 6) and the RWD context (16% for CDM).

Note that in this simulation, treatment, *A,* does not affect the outcome, *Y*. If treatment were to affect the outcome, then the true causal risk difference (CRD) would be different for different values of *B*, even if there were no interaction term between *B* and *A*, due to the properties of the $logit^{-1}$ function. Instead, because *A* does not affect *Y*, the true value of the CRD for all combinations of RCT and RWD is zero. This sets up an even competition between the study designs when different RWD are included in the hybrid analysis; we would expect the power to reject the same null hypothesis to depend on bias and variance but not on different true causal effects when different combinations of data are analyzed. We report 95% CI coverage for the true causal risk difference of zero for all designs across the 1000 iterations of this simulation.

We also aim to evaluate the amount of person-time that participants are precluded from receiving a GLP1-RA because they are in the control arm of one of the potential RCTs. As shown in Supplementary Table 1, for **Designs 1 and 3**, we start by determining whether the results of RCT1 or of the hybrid RCT1-RWD analysis reject the null hypothesis. Because the simulated effect is zero, and we expect a truly negative effect of semaglutide versus standard-of-care on MACE based on the results of the SUSTAIN 6 trial[16], we shift the definition of the null hypothesis to be that the risk difference is a specified value larger than zero.

For the sake of this demonstration, we consider a significant result as an estimate of the risk difference with an upper 95% CI limit less than positive 1.1%. This value was chosen because a trial of 9500 participants (similar to our simulated RCT2) would be expected to have power of 0.8 to detect a risk difference of -1.1% (using $\alpha = 0.05$). If we view the simulated data as having simulated CRD values that are shifted 1.1% more positive than the value of the CRD that our superiority trial would be powered to detect, then shifting standard criteria for superiority by the same amount would cause us to conclude that a result was significant if the estimated upper bound on the 95% confidence interval were less than positive 1.1%. Power is then calculated as the proportion of iterations in which this modified null hypothesis is rejected (i.e., the proportion of iterations in which the 95% CI did not include 1.1%).

Note that this method of evaluating significance should actually be slightly conservative given that in the simulated data, the probability of the outcome is approximately the PIONEER6 placebo arm rate (4.2%) in both arms, whereas if a negative risk difference had been simulated, the treatment arm outcome probability would have been less than 4.2%. For example, with the same sample size $N$, if the treatment arm probability of the outcome were 3%, the variance of the difference in sample proportions ($V1$) would be smaller than the variance if the treatment arm probability of the outcome were also 4.2% ($V2$):

$$V1 \approx \frac{0.03(1 - 0.03)}{N} + \frac{0.042(1 - 0.042)}{N} = \frac{0.069}{N}$$

$$< \frac{0.080}{N} = \frac{0.042(1-0.042)}{N} + \frac{0.042(1-0.042)}{N} \approx V2.$$

Finally, the person-time participants are prevented from receiving any GLP1-RA for each design in a single iteration of the simulation is calculated as described in Supplementary Table 1 below. We report the average amount of person-time during which participants were prevented from receiving a GLP1-RA across all iterations for each design. While the event-driven SOUL trial will actually run for closer to four years, we only include person-time required to evaluate the outcome for this proposed study: MACE by one year after baseline.

**Supplementary Table 1: Calculation of Person-Time prevented from Receiving a GLP1-RA for each Design**

| Design | Calculation of person-time prevented from receiving a GLP1-RA |
|---|---|
| 1 | 1. If RCT1 result is significant: 1 year x 1591.5 placebo arm participants = 1591.5 person-years<br>2. If RCT1 result not significant:<br>    a. 1 year x 1591.5 placebo arm participants from RCT1 + 1 year x 4750 placebo arm participants from RCT2 = 6341.5 person-years |
| 2 | 1. 1 year x 4750 placebo arm participants from RCT2 = 4750 person-years |
| 3 | 1. If hybrid RCT1-RWD result is significant[§]: 1 year x 1591.5 placebo arm participants from RCT 1 = 1591.5 person-years<br>2. If hybrid RCT1-RWD result not significant:<br>    a. 1 year x 1591.5 placebo arm participants from RCT1 + 1 year x 4750 placebo arm participants from RCT2 = 6341.5 person-years |

[§]RWD participants were not prevented from receiving a GLP1-RA by being in an RCT control arm and so are not included in the amount of person-time during which patients

are prevented from receiving a GLP1-RA.

A simple Super Learner[17] library of candidate algorithms was used for the ES-CVTMLE to improve computational efficiency of this simulation. The outcome regression was estimated using logistic regression. The Super Learner for the censoring mechanism ($P(C = 0|A, W)$) and treatment mechanism ($P(A = 1|W)$) for the combined RCT and RWD considered either a logistic regression or the sample mean. Because censoring was negligible in the RCT, the censoring mechanism in the RCT used within the ES-CVTMLE estimator only considered the sample proportion of non-missing outcomes. R Statistical Software version 4.2.2 was used for all simulations[18].

**Appendix 5: Data generation and results for simulation in which bias has no effect on NCO**

We also include a simulation in which the bias term, *B*, has no effect on the NCO. This simulation is included as a worst-case scenario for how hybrid **Design 3** could perform under a complete violation of the assumption that the factors causing bias in the relationship between the treatment and the true outcome also cause bias in the relationship between the treatment and the NCO. The process for generating the data for this simulation is the same as in Appendix 4, except that

$$NCO \sim Bernoulli\left(p = logit^{-1}(-3.33 + 0.2 * W1 - 0.4 * W2 + U_{nco})\right).$$

Supplementary Figure 1 shows the results of 1000 iterations of this simulation both when the ES-CVTMLE uses the estimated average treatment effect on the NCO as an estimate of the causal gap and when the ES-CVTMLE only estimates bias based on the method described in Appendix 3, without considering the NCO.

**Supplementary Figure 1: Simulation Results by Study Design with Different Amounts of RWD Bias when Bias has No Effect on NCO**

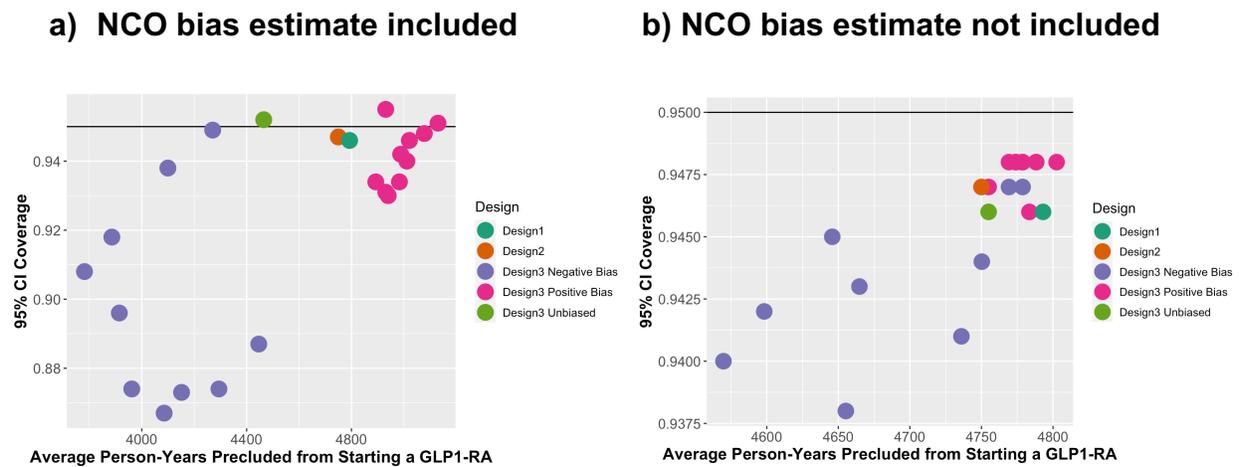

**Caption:** CI: Confidence Interval

Because the bias term, *B*, has no effect on the NCO in this simulation, hybrid **Design 3** is more likely to incorporate biased RWD in either direction when the NCO is used to estimate bias. With bias in the positive direction (towards the null), the hybrid RCT-RWD design is less likely to reject the null hypothesis, leading to the follow-up RCT being run in more iterations. As a result, patient-time during which participants were precluded from receiving a GLP1-RA was larger on average for **Design 3** with positive bias than for **Designs 1 and 2**.

With bias in the negative direction, **Design 3** led to as large as an average of 1012 fewer person-years during which participants could not start a GLP1-RA compared to **Design 1** but had 95% CI coverage ranging from 0.867 to 0.949. These results demonstrate the importance of choosing a good negative control outcome if this particular study design and estimator are selected.

If an appropriate negative control outcome is not available, the ES-CVTMLE may also only use one estimate of bias, as described in Appendix 3. In this context, the ES-CVTMLE is more conservative (less likely to include RWD), with coverage not less than 0.938 for any magnitude of bias, but a maximum average decrease in patient-years during which a GLP1-RA may not be prescribed of 223. The possibilities described in Supplementary Figure 1 should also be considered in the process of study design and estimator selection. Also note that the performance of **Design 1** compared to **Design 2** is slightly different in this simulation compared to the simulation in the main text. Such variability is expected when the same simulation is run different times although may be decreased by running more iterations.

**Appendix 6: Further details regarding specification of the ES-CVTMLE and unadjusted estimators**

The ES-CVTMLE estimator for the real data analysis used twenty cross-validation folds. The Super Learner libraries for the relevant regressions consisted of logistic regression or the sample mean for the outcome, logistic regression[19], a general additive model[20], or multivariate adaptive regression splines[21] for the propensity score, logistic regression[19], a general additive model[20], multivariate adaptive regression splines[21] or the sample mean for the RWD outcome censoring mechanism, and the sample mean only for the RCT outcome censoring mechanism. Because estimates vary somewhat when different random seeds are used to define cross-validation folds and train machine learning algorithms, the ES-CVTMLE estimator was run ten times with different random seeds, and the point estimate and upper and lower confidence interval bounds were averaged across these ten iterations.

Because censoring was negligible (0.3%) in the simulated and real data RCTs, a complete case analysis was conducted for the components of each study design that only involved RCT data.

**Appendix 7: Details for the construction of the CDM cohort**

Supplementary Tables 2-7 in this section describe how we translated the PIONEER 6 protocol into a format that could be implemented using the RWD. To minimize positivity issues, the cohort was trimmed to only include individuals with values of age, eGFR, HbA1c, HDL and LDL that fell within the 2.5-97.5% range of the values observed in PIONEER 6.

**Supplementary Table 2. Baseline Characteristics**

|  |  | **Comments** |
|---|---|---|
|  | **Age** | Continuous |
|  | **Sex** | Binary |
|  | **Race** | To avoid positivity issues resulting from many small categories of baseline confounding variables, we categorized race as:<br><br>1. White<br>2. Black<br>3. Other<br><br>However, the "other" group is quite different between PIONEER 6 (more Asian participants) and CDM (more Hispanic participants). Asian participants in PIONEER 6 were more likely to live in Asia, whereas Asian participants in CDM were living in the United States. |
| **Laboratory Measurements** |  | We used the most recent measurement within six months prior to time zero. If there was more than one measurement on that day, we took an average of those measurements. We also filtered out measurements outside trustworthy ranges based on standard data cleaning rules. |
|  | **HbA1c** | Continuous measurement |
|  | **HDL** | Continuous measurement |
|  | **LDL** | Continuous measurement |
|  | **eGFR** | We used the MDRD calculation in CDM, but PIONEER 6 uses CKD-epi for calculating eGFR. |

| **Medical history** | | Recorded any time prior to time zero, in any diagnosis position, and in either the in- or out-patient setting | |
|---|---|---|---|
| | **Prior MI** | **ICD9** Acute MI: 410.xx, Old MI: 412.xx | **ICD10** MI: I21.X, I22.X  Old MI: I25.2 |
| | **Prior stroke or TIA** | **ICD9** 431, 433.x1, 434.x1, 435.xx | **ICD10** Stroke: I61.X, I63.X,  Old stroke: I69.1, I69.3, G45.X |
| | **Prior heart failure** | **ICD9** 428.x, 402.01, 402.11, 402.91, 404.01, 404.11, 404.91, 404.03, 404.11, 404.13, 404.91, 404.93 | **ICD10** I11.0, I13.0, I13.2, I50.X |
| **Drug history** | | Any claim with the specific medication in the six months prior to time zero | |
| | **Cardiovascular medication** | CV medication included the following subgroups: antihypertensives, lipid-lowering, anti-thrombosis, diuretics | |
| | **T2DM medication** | A combination of any of the following: metformin, sulfonylureas (SU), thiazolidinediones (TZD), sodium-glucose transport protein 2 inhibitors (SGLT2-i). | |
| | **Insulin** | Any insulin | |
| **Other** | | | |

|  | Morbid obesity | This is based on ICD-9 and ICD-10 codes as there are only a limited number of BMI measurements in CDM. ICD9: 278.01, ICD10: E66.01, E66.2 |
|---|---|---|

**Supplementary Table 3. AHFS codes for medications**

| Grouping for drug history | Detailed grouping | AHFS |
|---|---|---|
| **Insulin** | All insulin | 68:20.08 |
|  | Long acting (including intermediate acting, human insulin and premix) | 68:20.08.08, 68:20.08.12, 68:20.08.16 |
|  | Short acting | 68:20.08.04 |
| **T2DM medication excluding insulin** | All T2DM medications excluding insulin | 68:20 but without: 68:20.08 (Insulins) |
|  | Metformin | 68:20.04 |
|  | TZD (rosiglitazone, pioglitazone) | 68:20.28 |
|  | SU | 68:20.20 |
| **CV medication** |  |  |

| | Antithrombotic (antiplatelet and anticoagulants) | 20:12.04.08 |
| | | 20:12.04.12 |
| | | 20:12.04.14 |
| | | 20:12.18 |
| | | 20:12.20 |
| | Antihypertensives | 24:08 (minus diuretics 24:08.24) |
| | | 24:12 |
| | | 24:20 |
| | | 24:24 |
| | | 24:28 |
| | | 24:32 |
| | Diuretics (pooled with antihypertensives) | 24:08.24 |
| | Lipid lowering | 24:06.04, |
| | | 24:06.05, |
| | | 24:06.06, |
| | | 24:06.08, |
| | | 24:06.24 |

**Supplementary Table 4. Primary MACE outcome in the PIONEER 6 trial and the translation to CDM**

| **PIONEER-6 trial definitions** | **ICD-9** | **ICD-10** |

| First occurrence of<br>· Nonfatal myocardial infarction (MI)<br>· Nonfatal stroke<br>· Death from all cause | First occurrence measured in diagnosis position 1 and inpatient care setting | |
|---|---|---|
| | **MI**: 410.X<br>**Stroke**: 431.X, 433.X1, 434.X1<br>**Mortality**: All cause death | **MI**: I21.X, I22.X<br>**Stroke**: I61.X, I63.X<br>**Mortality**: All cause death |

**Supplementary Table 5. Inclusion criteria in the PIONEER 6 trial and the translation to CDM**

| PIONEER 6 inclusion criteria | ICD9 | ICD 10 |
|---|---|---|
| Informed consent | **N/A** | |
| Men or women with type 2 diabetes | Our CDM cohort only includes individuals with diabetes.<br><br>Excluded patients with type 1 diabetes (T1D):<br>Measured at any time in any diagnosis position and inpatient or outpatient care setting:<br>T1D (ICD-9 250.x1, ICD-9 250.x3) | E10.X, O24.0 |
| Either of the following: | | |

| Age ≥50 years at screening and at least one of the following conditions: | Age ≥ 50 at time zero AND at least one of the following measured prior to time zero in any diagnosis position and inpatient or outpatient care setting: | |
|---|---|---|
| Prior MI | Acute MI: 410.xx, Old MI: 412.xx | MI: I21.X, I22.X<br>Old MI: I25.2 |
| Prior stroke or TIA | Stroke or TIA: 431, 433.x1, 434.x1, 435.xx | Stroke: I61.X, I63.X,<br>Old stroke: I69.1, I69.3<br>TIA G45.X |
| Prior coronary, carotid or peripheral arterial revascularization including:<br>Percutaneous transluminal coronary angioplasty (PTCA)<br>Coronary artery bypass graft (CABG) | Coronary revascularization (PTCA, stenting, CABG)<br>PTCA:<br>ICD-9 proc: 00.66, 36.01, 36.02, 36.03, 36.05, 36.09<br>Stenting:<br>ICD-9 proc: 36.06, 36.07<br>CABG:<br>ICD-9 proc: 36.1x, 36.2<br>Transmyocardial revascularization:<br>ICD-9 proc: 36.31-36.34 | ICD-10-PCS: 027.X, 021.X |
| >50% stenosis on angiography or imaging of coronary, carotid, or lower extremity arteries | Other forms of chronic ischemic heart disease<br><br>414.x<br><br>Peripheral vascular disease:<br><br>440.20 – 440.24, 440.29 – 440.32, 440.3, 440.4, 443.9 | Chronic ischemic heart disease: I25.X<br><br>Peripheral vascular disease: I70.X, I73.9 |
| History of symptomatic coronary heart disease documented by positive exercise stress test or any cardiac imaging or unstable angina with ECG changes | N/A | |
| Asymptomatic cardiac ischemia documented by positive nuclear imaging test, exercise test or dobutamine stress echo | N/A | |
| Chronic heart failure NYHA class II-III | Heart Failure: 428.x, 402.01, 402.11, 402.91, 404.01, 404.11, 404.91, 404.03, 404.11, 404.13, 404.91, 404.93 | I11.0, I13.0, I13.2, I50.X |
| Moderate renal impairment (estimated glomerular filtration rate [eGFR] 30 to 59 ml/min/1.73 m$^2$) | Chronic Kidney Disease (CKD) stage 3:<br><br>585.3 | CKD stage 3 N18.3 |
| **Age ≥60 years at screening and at least one of the following risk factors:** | Age ≥ 60 at time zero AND measured prior to time zero in any diagnosis position and inpatient or outpatient care setting: | |
| Microalbuminuria or proteinuria | Proteinuria: 791.0 Albumin abnormality: 790.99 | Proteinuria: R18.X Albumin abnormality: R77.0 |
| Hypertension and left ventricular hypertrophy by ECG or imaging | N/A | |
| Left ventricular systolic or diastolic dysfunction by imaging | N/A | |
| Ankle-brachial index <0.9 | Atherosclerosis of native arteries of the extremities with intermittent claudication: 440.21 | Atherosclerosis of native arteries of the extremities with intermittent claudication: I70.21 |

**Supplementary Table 6. Exclusion criteria in the PIONEER 6 trial and the translation to CDM**

| PIONEER 6 exclusion criteria | ICD-9 | ICD-10 |
|---|---|---|
| Known or suspected hypersensitivity to the trial product or related products. | N/A | |
| Previous participation in this trial. Participation is defined as signed informed consent | N/A | |

| | |
|---|---|
| Females of childbearing potential who are pregnant, breast-feeding or intend to become pregnant or are not using adequate contraceptive methods (adequate contraceptive measures as required by local law or practice) | N/A |
| Receipt of any investigational medicinal product within 90 days before screening. | N/A |
| Participation in another clinical trial of an investigational medicinal product. Participation in a clinical trial which evaluates stent(s) is allowed. | N/A |
| Current or previous (within 90 days prior to screening) treatment with any GLP-1 receptor agonist, DPP-4 inhibitor or pramlintide | **Dispensing of at least one of the following medications in the 90 days prior to index date:**<br>Use of a GLP-1 receptor agonist ("exenatide", "liraglutide", "lixisenatide", "albiglutide", "dulaglutide", "semaglutide", "beinaglutide")<br>**or**<br>pramlintide<br>**or**<br>any dipeptidyl peptidase 4 (DPP-4) inhibitor ("sitagliptin", "vildagliptin", "saxagliptin", "alogliptin", "linagliptin", "gemigliptin", "evogliptin", "teneligliptin") |
| Any disorder, which in the investigator's opinion might jeopardize subject's safety or compliance with the protocol. | N/A |

| | | |
|---|---|---|
| Family or personal history of multiple endocrine neoplasia type 2 (MEN 2) or medullary thyroid carcinoma (MTC) | Measured any time prior to time zero in any diagnosis position and inpatient or outpatient care setting:<br><br>MEN Type IIA:<br><br>258.02<br><br>MEN Type IIB:<br><br>258.03 | MEN Type IIA: E31.22<br>MEN, Type IIB: E31.23 |
| History of pancreatitis (acute or chronic). | Measured any time prior to time zero in any diagnosis position and inpatient or outpatient care setting:<br><br>Acute pancreatitis:<br><br>577.0<br><br>Chronic pancreatitis:<br><br>577.1 | K85, K86.0, K86.1 |
| History of major surgical procedures involving the stomach potentially affecting absorption of trial product (e.g., subtotal and total gastrectomy, sleeve gastrectomy, gastric bypass surgery). | Measured any time prior to time zero in any diagnosis position and inpatient or outpatient care setting:<br><br>Partial gastrectomy:<br><br>43.5x-43.8x<br><br>Total gastrectomy:<br><br>43.9x<br><br>Sleeve gastrectomy:<br><br>43.82<br><br>Gastric bypass:<br><br>44.3x,44.68, 44.95, 44.96, 44.97, 44.99, 44.5 | Bypass of stomach 0D16<br><br>Excision/resection of stomach 0DB6, 0DT6 |
| Subjects presently classified as being in New York Heart Association (NYHA) Class IV heart failure. | N/A | |
| Planned coronary, carotid or peripheral artery revascularization known on the day of screening | N/A | |
| Any of the following: myocardial infarction, stroke or hospitalization for unstable angina or transient ischemic attack within the past 60 days prior to screening. | Measured 60 days prior to time zero in any diagnosis position and inpatient or outpatient care setting:<br><br>MI, Stroke<br>Acute MI:<br>410.xx<br>Stroke:<br>431, 433.x1, 434.x1<br><br>Measured 60 days prior to time zero in any diagnosis position and inpatient care setting:<br>TIA:<br>435.xx<br><br>Unstable angina:<br>ICD9: 411.1 | **MI:** I21.X, I22.X<br>**Stroke:** I61.X, I63.X<br><br>Unstable angina: ICD10: I20.0, I25.110, I25.700, I25.710, I25.720<br><br>TIA: G45 |
| Chronic or intermittent hemodialysis or peritoneal dialysis or severe renal impairment (corresponding to eGFR <30 mL/min/1.73 m2). | Measured any time prior to time zero in any diagnosis position and inpatient or outpatient care setting:<br><br>CKD stage 4:<br><br>585.4<br><br>CKD stage 5:<br><br>585.5<br><br>ESRD:<br><br>585.6<br><br>Measured any time prior to time zero in any diagnosis position and inpatient or outpatient care setting at least twice:<br><br>Hemodialysis:<br><br>ICD9Proc: 39.95<br><br>Peritoneal dialysis:<br><br>ICD9Proc: 54.98 | CKD stage 4: N18.4<br>CKD stage 5: N18.5<br>ESRD: N18.6<br><br><br>Hemodialysis: 5A1D<br><br>Peritoneal dialysis 3E1M<br><br>Trimming handles eGFR already. |
| History or presence of malignant neoplasms within the last 5 years (except basal and squamous cell skin cancer and carcinoma in situ). | Measured at any time prior to time zero in any diagnosis position and inpatient or outpatient care setting<br><br>History of malignant neoplasm:<br><br>140.xx-208.xx (except 173.xx, non-melanoma skin cancer) | C01-C99 except C44 |
| History of diabetic ketoacidosis. | Measured at any time prior to time zero in any diagnosis position and inpatient or outpatient care setting:<br><br>Secondary diabetes mellitus with ketoacidosis:<br><br>249.1<br><br>Diabetes with ketoacidosis: | E08.1<br>E11.1<br>E13.1 |

| | 250.1 | |
|---|---|---|
| Proliferative retinopathy or maculopathy requiring acute treatment. Verified by fundus photography or dilated fundoscopy performed within 90 days prior to screening or within the period between screening and randomization. | N/A (due to difficulty in identifying acute treatment) | |

**Supplementary Table 7. Negative control outcome in the PIONEER 6 trial and the translation to CDM**

| PIONEER-6 trial definitions | ICD-9 | ICD-10 |
|---|---|---|
| First occurrence of fracture | First occurrence measured in diagnosis position 1 and inpatient care setting ||
| | 800-829 | S02, S12, S22, S32, S42, S52, S62, S72, S82, S92 |
|   |   |   |

## Appendix 8: References for Appendices